# Atmospheric modeling: Setting Biomarkers in context

L. Kaltenegger

*Harvard-Smithsonian Center for Astrophysics, 60 Garden Street, Cambridge, MA 02138, USA*

F. Selsis

*LAB: Laboratoire d'Astrophysique de Bordeaux (CNRS; Université Bordeaux I), BP 89, F-33270 Floirac, France*

**Motivation:**
ESA's goal to '*detect biomarkers in Earth-like exoplanets in the Habitable Zone*' requires theoretical groundwork that needs to be done to model the influence of different parameters on the detectable biomarkers.

*We need to model a wide parameter space (chemical composition, pressure, evolution, interior structure and outgassing, clouds) to generate a grid of models that inform our detection strategy as well as can help characterize the spectra of the small rocky planets detected.*

**Abstract**
To understand the spectrum of a detected exoplanet, we need to set it in context. The spectrum of the planet can contain signatures of atmospheric species that are important for habitability, like $CO_2$ and $H_2O$, or resulting from biological activity ($O_3$, $CH_4$, and $N_2O$). The presence or absence of these spectral features will indicate similarities or differences with the atmospheres of terrestrial planets and are discussed in detail and set into context with the physical characteristics of a planet in this chapter. A crucial step in this process is the theoretical groundwork that needs to be done to model the influence of different parameters on the detectable biomarkers.

In a famous paper, Sagan et al. (1993) analyzed a spectrum of the Earth taken by the Galileo probe, searching for signatures of life. They concluded that the large amount of $O_2$ and the simultaneous presence of $CH_4$ traces are strongly suggestive of biology. The detection of a widespread red-absorbing pigment with no likely mineral origin supports the hypothesis of biophotosynthesis. The search for signs of life on possibly very different planets implies that we need to gather as much information as possible in order to understand how the observed atmosphere physically and chemically works. We need to model a wide parameter space (chemical composition, pressure, evolution) to generate a grid of models that inform our detection strategy as well as can help characterize the spectra of the small rocky planets detected.

## Introduction

Over 300 giant exoplanets already have been detected, and hundreds, perhaps thousands more, are anticipated in the coming years. The nature of these planets, including their orbits, masses, sizes, constituents, and likelihood that life could develop on them, can be probed by a combination of observations and modeling. So far, our detection methods from the ground, Radial Velocity and Transit Search, are biased towards big planets orbiting close to their parent star because they are easier to detect. Those planets produce a bigger, more frequent signal than a small planet further away from its parent star. Even so, the number of smaller EGP found, indicates a trend towards smaller masses. The detection of the first potentially rocky planets have already been announced, SuperEarths.

## Biomarkers

Biomarkers (or biosignature) is used here to mean detectable species, or set of species, whose presence at significant abundance strongly suggests a biological origin. This is for instance the case for the couple $CH_4+O_2$. Bio-indicators are indicative of biological processes but can also be produced abiotically in significant quantities. Our search for signs of life is based on the assumption that extraterrestrial life shares fundamental characteristics with life on Earth, in that it requires liquid water as a solvent and has a carbon-based chemistry. Life on the base of a different chemistry is not considered here because the vast possible life-forms produce signatures in their atmosphere that are so far unknown.

The first step of exploration in terrestrial extrasolar planets will be a space mission that can detect and record low resolution spectra of extrasolar planets like ESA's proposed Darwin mission. $O_2$, $O_3$, $CH_4$ are good biomarker candidates that can be detected by a low-resolution (Resolution < 50) spectrograph. There are good biogeochemical and thermodynamic reasons for believing that these gases should be ubiquitous byproducts of carbon-based biochemistry, even if the details of alien biochemistry are significantly different than the biochemistry on Earth.

We need to model and understand the abiotic sources of biomarkers better, so that we can identify when it might constitute a "false positive" for life detection, when abiotic sources could produce high quantities of a species we understand as a biomarker on Earth. Other biogenic trace gases might also produce detectable biosignatures e.g. $N_2O$. Currently identified potential candidates include volatile methylated (like $CH_3Cl$) and sulfur compounds. Although it is known that these compounds are produced by microbes, it is not yet fully understood how stable (or detectable) these compounds would be in atmospheres of different composition and for stars of different spectral type and incident UV flux. These uncertainties, however, could be addressed by further modeling studies. If we can expand the potential suite of detectable biogenic trace gases, and understand the condition under which they are most likely to be detectable, we will gain more confidence in our ability to identify life remotely.

*The theoretical modeling research goals are to explore the plausible range of habitable planets and to improve our understanding of the detectable ways in which life modifies a planet on a global scale.*

## Modeling a habitable Planet

On an Earth-like planet where the carbonate-silicate cycle is at work, the level of $CO_2$ in the atmosphere depends on the orbital distance: $CO_2$ is a trace gas close to the inner edge of the HZ but a major compound in the outer part of the Habitable Zone (HZ). Earth-like planets close to the inner edge are expected to have a water-rich atmosphere or to have lost their water reservoir to space. The limits of the HZ are known qualitatively, more than quantitatively. This uncertainty is mainly due to the complex role of clouds but also to three-dimensional climatic effects not yet included in the modeling. Thus, planets slightly outside the computed HZ could still be habitable, while planets at habitable orbital distance may not be habitable because of their size or chemical composition.

The range of characteristics of planets is likely to exceed our experience with the planets and satellites in our own Solar System as first detections of exoplanets show. Earth-like planets orbiting stars of different spectral type might evolve differently. Models of such planets need to be generated, considering the changing atmosphere structure, clouds, as well as the interior structure of the planet. One crucial factor in interpreting planetary spectra is the point in the evolution of the atmosphere when its biomarkers become detectable.

*Those spectra will be used as part of a big grid that should be generated to characterize any exoplanets found. This also influences the design requirements for a spectrometer.*

The search for signs of life implies that one needs to gather as much information as possible in order to understand how the observed atmosphere physically and chemically works. Knowledge of the temperature and planetary radius is crucial for the general understanding of the physical and chemical processes occurring on the planet (tectonics, hydrogen loss to space). This question is far from being easy. In theory, spectroscopy can provide some detailed information on the thermal profile of a planetary atmosphere. This however requires a spectral resolution and a sensitivity that are well beyond the performance of a first generation spacecraft like Darwin/TPF. Thus we will concentrate on the initially available observations here.

**Model Temperature and Radius of the Planets**

One can calculate the stellar energy of the star that is received at the measured orbital distance. This only gives very little information on the temperature of the planet which depends on its albedo. The surface temperature is likely to be enhanced by greenhouse gases – therefore it is important to observe the spectral signatures of the greenhouse gases for the atmosphere models. However, with a low resolution spectrum of the thermal emission, the mean effective temperature and the radius of the planet can be obtained by fitting the envelope of the thermal emission by a Planck function. The ability to associate a brightness temperature to the spectrum relies on the existence and identification of spectral windows probing the surface or the same atmospheric levels. Such identification is not trivial in the absence of any other information on the observed planet and needs to be modeled in detail.

**Summary**

Any information we collect on habitability, is only important in a context, that allows us to interpret, what we find. Knowledge of the temperature and planetary radius is crucial for the general understanding of the physical and chemical processes occurring on the planet. These parameters as well as an indication of habitability can be determined with low resolution spectroscopy and low photon flux, as assumed for first generation space missions.

*We need to model a wide parameter space (chemical composition, pressure, evolution, interior structure and outgassing, clouds) to generate a grid of models that inform our detection strategy as well as can help characterize the spectra of the small rocky planets detected.*

An amazing feature of future space based missions like Darwin and TPFs is that they will have the capability to, for the first time, do comparative planetology on a wide variety of planets, whose atmosphere and climatology is far outside our current understanding, as well as find planets similar to our own and probe them for habitable conditions.

**References**
1 Sagan, C., Thompson, W. R., Carlson, R., Gurnett, D., and Hord, C.: 1993, *Nature*, **365**, 715.
2 Kasting, J.F., Whitmire, D.P., and Reynolds, H.: 1993, *Icarus*, **101**, 108.
3 Des Marais, D. J., Harwit, M. O., Jucks, K. W., Kasting, J. F., Lin, D. N. C., Lunine, J. I., Schneider, J., Seager, S., Traub, W.A., Woolf, N. J.: 2002, *Astrobiology*, **2**, 153.
4 Owen, T.: 1980, *Strategies for Search for Life in the Universe*, Dordrecht, The Netherlands, 177.
5 Lovelock J.E., *Proc.R. Soc. Lond. B*, **189**, 167.
6 Paillet, J.: 2006, Spectral characterization of terrestrial exoplanets, *phD thesis*.
7 Traub, W.A. and Jucks, K.: 2002, *AGU Monograph Series*, **130**, 369.
8 Leger, A., Pirre, M., and Marceau, F. J.: 1993, *A&A*, **277**, 309.
9 Beichman, C. A., Woolf, N. J., and Lindensmith, C. A.: 1999, The Terrestrial Planet Finder (TPF): a NASA Origins Program to Search for Habitable Planets / the TPF ScienceWorking Group, Washington DC: NASA JPL, *JPL publication*, **99–3**.
10 Beichman, C.A., Fridlund, M., Traub, W.A., Stapelfeldt, K.R., Quirrenbach, A., and Seager, S.: 2006, *Protostars and Planets V*, Tuscon: University of Arizona Press.
11 Fridlund, M.: 2000, DARWIN The InfraRed Space Interferometer, (Leiden: ESA), ESA-SCI(2000), **12**, 47.
12 Kaltenegger, L.: 2004, Search for Extraterrestrial Planets: The DARWIN mission. Target Stars and Array Architectures, *phD thesis astro-ph/0504497*.
13 Kaltenegger, L. and Fridlund, M.: 2005, *Advances in Space Research*, **36**, 1114.
14 Segura, A., Krelove, K., Kasting, J. F., Sommerlatt, D., Meadows V., Crisp D., Cohen, M., and Mlawer, E.: 2003, *Astrobiology*, **3**, 689.
15 Selsis, F.: 2000, Darwin and astronomy: the infrared space interferometer, *ESA SP*, **451**, 133.
16 Segura, A., Kasting, J.F., Meadows, V., Cohen, M., Scalo, J., Crisp, D., Butler, R. A. H., and Tinetti, G.: 2005, *Astrobiology*, 706.
17 Valencia, D., et al.: 2006, *Icarus*, **181**.
18 Schindler, T. L. and Kasting, J. F.: 2000, *Icarus*, **145**, 262.
19 Pavlov, A.A., Kasting, J.F., Brown, L.L., Rages, K.A., Freedman R., and Greenhouse R.: 2000, *J. Geophys. Res.*, **105**, 981.
20 Kaltenegger, L., Traub, W.A., and Jucks, K. 2007, *ApJ*.658, 598
21 Arnold, L., Gillet, S., Lardiere, O., Riaud, P., and Schneider, J.: 2002, *A&A*, **392**, 231.
22 Montanes-Rodriguez P., Palle E., and Goode P.R.: 2006, *ApJ*.
23 Montanes-Rodriguez P., et al.: 2007, *ApJ*, **629**.
24 Seager, S. and Ford, E.B.: 2002, Astrophysics of Life, Cambridge: Cambridge University Press, 57.
25 Selsis, F., Despois, D., Parisot, J.-P.: 2002, *A&A*, **388**, 985.
26 Kasting, J.F.: 1997, *Origins of Life*, 291.
27 Forget, P. and Pierehumbert, H.: 1997, *Science*, **278**.
28 Walker, J.C.G.: 1977, *Evolution of the Atmosphere*, Macmillan, New York.
29 Christensen, P.R., Pearl, J.C.: 1997, *J. Geoph. Res.*, **102**, 875.
30 Woolf, N. J., Smith, P. S., Traub, W. A., and Jucks, K. W.: 2002, *ApJ*, **574**, 430.
31 Kasting, J.F. and Catling, D.: 2003, *Ann.Rev. Astron. Astrophys.*, **41**, 429.
32 Kasting, J.F.: 2004, *Scientific American*, **80**.
33 Seager, S., Turner, E. L., Schafer, J., and Ford, E. B.: 2005, *Astrobiology*.
34 Kaltenegger, L., Eiroa, C., and Fridlund, M.: 2007, *A&SS in press*
35 Selsis, F.: 2003, *Proc. ESA-SP539, ESA, Netherlands*.
36 Gaidos, E. and Williams, D. M.: 2004, *New Astronomy*, **10**, 67.
37 Kaltenegger, L., and Selsis, F. 2007 Biomarkers set in context, in Extrasolar Planets, Wiley-VCH, Textbook Physics
38. Tinetti, G., Meadows V. S., Crisp D., Fong W., Fishbein E., Turnbull M., and Bibring J.-P. 2006, Astrobiology 6, 34
39 Selsis, F; Kaltenegger, L; Paillet, J. 2008, Physica Scripta, Volume 130, Issue , pp. 014032